\documentclass[aps,prl,twocolumn,superscriptaddress,showpacs,floatfix,longbibliography]{revtex4-2}
\usepackage{amsmath,amssymb,amsfonts,float,graphics,epsfig,epstopdf,color,verbatim,tabularx,bm,multirow,appendix,hyperref}
\usepackage[normalem]{ulem}
\usepackage{lmodern}
\usepackage{ytableau}
\usepackage{pifont}
\usepackage{color}
\usepackage{bm}

\begin{document}
	
\title{The ``Sign problem" of the 3rd order anomalous Hall effect \\ in topological magnetic materials}
\author{Xu Zhang}
\affiliation{Department of Physics and HKU-UCAS Joint Institute of Theoretical and Computational Physics, The University of Hong Kong, Pokfulam Road, Hong Kong SAR, China}
\author{Kai Sun}
\email{sunkai@umich.edu}
\affiliation{Department of Physics, University of Michigan, Ann Arbor, Michigan 48109, USA}
\author{Zi Yang Meng}
\email{zymeng@hku.hk}
\affiliation{Department of Physics and HKU-UCAS Joint Institute of Theoretical and Computational Physics, The University of Hong Kong, Pokfulam Road, Hong Kong SAR, China}

\begin{abstract}
The anomalous Hall effect (AHE), which provides a bridgeway between the geometry of quantum wavefunctions and transport measurements, has been a key focus of intensive studies.
In addition to the well-studied linear AHE, governed by the electronic Berry curvature, nonlinear AHE originated from higher-order Berry-curvature multipoles has also been observed in recent studies. 
Inspired by the 3rd order AHE and its room temperature sign switching in kagome antiferromagnet FeSn~\cite{berthold2023}, we investigate the generic sign structure of Berry-curvature-induced 3rd order AHE in topological magnetic material. We find that in contrast to the linear Hall coefficient, whose sign is determined by the broken time-reversal symmetry, the sign of the 3rd order Hall coefficient is dictated by the interplay between time-reversal symmetry breaking, magnetic order, and spin-orbit couplings. Our calculations give a possible solution for the ``sign problem" of the 3rd order AHE response in the phase space spanned by the in- and out-of-plane magnetization, the spin-orbital coupling strength and chemical potential. We further propose realistic experiment setups to systematically reveal the sign structure in the 3rd order AHE response
via continuously rotating the magnetic field directions.
\end{abstract}
\date{\today}
\maketitle

\noindent{\textcolor{blue}{\it Introduction.}---} 
The Hall effect -- the development of transverse Hall voltage in a material carrying a finite electrical current subjected to an out-of-plane magnetic field -- was discovered in the 19th century~\cite{Hall1879} and has by now evolved into the common knowledge of condensed matter physics. The anomalous Hall effect (AHE), on the other hand, describes the situation where the Hall effect appears in metallic ferromagnets independent of the external magnetic field and was discovered similar in time~\cite{Hall1881}, but the important and fundamental connection with topology and Berry phase of the quantum metric in electronic states was fully appreciated much later~\cite{berry1984quantal,haldaneBerry2004,Nagaosa2010,DiXiao2010}. Thanks to such connection, in recent years, the AHE leads to many exciting discoveries in topological quantum materials, such as the quantum AHE in ferromagnet topological insulators~\cite{changExperimental2013,dengQuantum2020}, room temperature large AHE in a non-collinear antiferromagnet~\cite{nakatsujiLarge2015}, AHE in Weyl and Dirac kagome metals~\cite{yeMassive2018,dzsaber2021giant,liuGiant2018,wangLarge2018,kumar2021room} and (quantum and nonlinear) AHE in quantum moir\'e materials~\cite{huang2022giant,sinha2022berry,chakraborty2022nonlinear,lai2021third,Benjamin2020,qin2021strain,kang2019nonlinear,ma2019observation,huang2022intrinsic,stepanovCompeting2021}.

Theoretically, it is well understood that the linear AHE response can originate from the electronic Berry curvature monopole~\cite{haldaneBerry2004,Nagaosa2010,DiXiao2010}. More recently, it is pointed out that the higher-order multipoles of Berry curvature can give rise to even richer nonlinear AHE responses, both in theoretical~\cite{Tonylow2015,Sodemann2015,Facio2018,You2018,Zhang_2018,YangZhang2018,Du2018,Parker2019,zhangHigher2020,Benjamin2020,Sobhit2020,DingFu2020,du2021nonlinear,ortix2021nonlinear,he2021giant,chuanchang2021,ChengPing2022,hu2022nonlinear,chakraborty2022nonlinear} and experimental~\cite{ma2019observation,kang2019nonlinear,shvetsov2019nonlinear,dzsaber2021giant,qin2021strain,ho2021hall,kumar2021room,lai2021third,huang2022intrinsic,huang2022giant,cao2022,sinha2022berry,berthold2023} activities of topological quamtum materials. 
The nonlinear AHE describes the phenomenon when an AC 
current with frequency $\omega$ is injected,  a voltage at integer multiples of the input frequency emerges along the orthogonal direction, and this signal is nonlinear with respect to the current and/or external fields. 

Experimental studies about the 2nd-order AHE are first carried out in non-centrosymmetric systems with time reversal symmetry, such as the strained twisted  graphene layers~\cite{ho2021hall,sinha2022berry,huang2022intrinsic} and transition metal dichalcogenides (TMD)~\cite{ma2019observation,kang2019nonlinear,shvetsov2019nonlinear,qin2021strain,huang2022giant}.
The 3rd-order AHE (and other odd-orders) requiring time-reversal symmetry breaking is observed in TMD and more recently, layered kagome antiferromagnet FeSn~\cite{lai2021third,berthold2023,salesElectronic2019,kangDirac2020}.  
The latter is of particular interests to the present paper, where the 1st- and 3rd- order Hall signals are observed from cryogenic conditions to above room temperature, but the 2nd-order response remains absent~\cite{berthold2023}. More remarkably,
near the magnetic transition temperature ($T_N = 365$ K~\cite{salesElectronic2019,xieSpin2021}), where spin canting 
induced by the magnetic field is the strongest, the 3rd order AHE signal switches sign while the sign of the 1st-order response remains unchanged~\cite{berthold2023}, indicating that the sign of 3rd-order Hall coefficient is very sensitive to the underlying magnetic and electronic properties, in direct contrast to its 1st- or 2nd-order counterpart. 
This observation reveals the rich sign structure of the 3rd order nonlinear AHE at room temperature and demonstrates its potential usage as a sensitive transport probe of magnetic properties in various topological materials.

Although highly intriguing, the physical origin of this rich and interesting sign structure of the 3rd order AHE remains yet to be understood, which will be refereed to as the ``sign problem" of the 3rd order AHE
~\footnote{The sign problem is usually refers to the  negative or complex sign of the configurational weigths in quantum Monte Carlo simulations for many-body systems, it is one of the NP-hard problem whose solution would results in a complete solution of many famous problems in science and technology, see recent review in Ref.~\cite{panSign2022}}.
The answer and insights to this ``sign problem", as well as the key physics principles that governs this sign structure,  is  of crucial importance for predicting this rich sign structure in the vast material and experiment related parameter space and guiding new investigations in higher-order AHE. 

This paper is our effort to give a possible solution for the ``sign problem". First, we review the higher order AHE from Berry curvature multipole according to semi-classical Boltzmann equation and derive the formula for the 3rd order AHE as simple weighted Fermi surfaces summation in isotropic system. Then we show, using a minimal model of topological magnetic materials, the generic requirements of the Berry curvature quadrupole induced 3rd order AHE and its rich sign structure in the phase diagram spanned by the in- and out-of-plane magnetization $m_x$ and $m_z$, the spin-orbital coupling strength $t_{soc}$ and chemical potential $\mu$. Finally, we further propose realistic experiment setup by continuously rotating the magnetic field directions to systematically reveal the highly sensitive and useful sign switching mechanism in the 3rd order AHE.


\noindent{\textcolor{blue}{\it Higher order AHE.}---} It is well-known that the anomalous velocity defined by Berry curvature can contribute to linear AHE conductance according to the semi-classical Boltzmann equation~\cite{DiXiao2010,Nagaosa2010}
\begin{eqnarray}
	\sigma_{\alpha\beta}=-q^2\epsilon_{\alpha\beta z} \sum_n \int d^2k f^{(0)}_n \Omega_n,
	\label{eq:eq1}
\end{eqnarray}
where $q$ is the charge of the carrier, $\epsilon_{\alpha\beta z}$ is the Levi-Civita symbol, $n$ is the band label, $\Omega$ indicates the Berry curvature along $z$ direction and $f^{(0)}$ is Fermi-Dirac distribution.

By recursively solving the semi-classical Boltzmann equation, the 3rd order AHE conductance contributed from higher momenta of Berry curvature can also be readily derived~\cite{Sodemann2015,Parker2019,zhangHigher2020} 
\begin{eqnarray}
	\sigma_{\alpha\beta\gamma}^{(2\omega)}&=&-\frac{q^3\tau}{1+i\omega\tau}\epsilon_{\alpha\beta z} \sum_n \int d^2k f^{(0)}_n \partial_{\gamma}\Omega_n, \nonumber\\
	\sigma_{\alpha\beta\gamma\mu}^{(3\omega)}&=&\frac{q^4\tau^2 \epsilon_{\alpha\beta z}}{(1+i2\omega\tau)(1+i\omega\tau)} \sum_n \int d^2k f^{(0)}_n \partial_{\gamma}\partial_{\mu}\Omega_n, \nonumber\\
\end{eqnarray}
where $\omega$ is the frequency of the driving electric field and $\tau$ is the relaxation time. 
Here we focus on the low frequency regime and disorder contributions are ignored.

In a typical band structure, the Berry curvature $\Omega$ is not uniformly distributed in the $k$-space. Instead, its intensity often peaks around some ``hotspots", generated by gapped Dirac or (other band) crossings~\cite{Haldane1988} . Therefore, to the leading order approximation, we can focus on these ``hotspots", which give dominant contributions to the Hall signal, and describe them using the $k\cdot p$ theory (an explicit example will be given below). For simplicity, here we assume that the dispersion and Berry curvature around such a ``hotspot" is isotropic (up to some rescaling of $k$). As pointed out by Haldane~\cite{haldaneBerry2004}, at low temperature and frequency, the Hall coefficients is fully dictated by electron wave functions near the Fermi surface. For nonlinear Hall response, following the same intuition, 
we rewrite $\sigma_{yxxx}^{(3\omega)} \propto \sum_n \int d^2k f^{(0)}_n \partial_{k_x}\partial_{k_x}\Omega_n$ (within the  approximation mentioned above) at $T=0$ as
\begin{equation}
\sigma_{yxxx}^{(3\omega)} \propto \sum_{n,s} sgn(\partial_{k} E_n) k \partial_{k} \Omega_n |_{k=k_{\mu,n,s}}
\label{eq:eq3}
\end{equation}
Here instead of integrating over the entire Fermi sea, this formula sums over Fermi surfaces, where $s$ labels the Fermi surfaces and $k_{\mu,n,s}$ is its Fermi wave vector. 
$E_n(k)$ is the dispersion of the $n$th band and $sgn(\partial_{k} E_n)=\pm1$ is the sign of the Fermi velocity $\partial_{k} E_n$ at this Fermi surface ($+/-$ for an electron/hole pocket). 

In a magnetic material, if we don't flip the orientation of magnetic moment, the sign of $\partial_{k} \Omega_n$ will usually remain the same, insensitive to temperature and other control parameters. Thus, if a material has only one pocket (with a fixed sign for $\partial_{k} E_n$), the sign of $\sigma_{yxxx}^{(3\omega)}$ is uniquely determined and thus cannot change. In contrast, for a system with two or more (electron/hole) pockets, contributions from different Fermi surfaces may carry opposite signs. In this picture, each Fermi surface contributes $\pm1$ with a weight of $|k \partial_{k} \Omega|$. Because control parameters (e.g. temperature or doping) can change the weight of each Fermi surface, the competition between $+$ and $-$ signs can lead to a rich sign structure for  $\sigma_{yxxx}^{(3\omega)}$.

It is also worthwhile to highlight that Eq.~\eqref{eq:eq3} and this sign-switching behavior can be naturally generalized to any odd higher order Hall coefficients, while the 3rd order AHE is just one example and the lowest channel to observe this phenomenon.

\noindent{\textcolor{blue}{\it Minimal model.}---} 
To illustrate the sign structure in 3rd order AHE, here we introduce a minimal model to describe the Berry curvature in a 2D magnetic material with spin-orbit coupling and magnetic moments. Near a Berry curvature hotspot, the $k\cdot p$ theory involves a 4-band Dirac Hamiltonian
\begin{align}
	H=\left(\begin{array}{cc} 
	 \mathbf{k}\cdot \boldsymbol{\sigma} +t_{soc} \sigma_z+m_z I  	& 	m_x I \\
	 m_x I	&	  \mathbf{k}\cdot \boldsymbol{\sigma} -t_{soc} \sigma_z	-m_z I
	\end{array}\right).
	\label{eq:eq4}
\end{align}
where $I$ and $\sigma$ are the identity and Pauli matrices respectively and $ \mathbf{k}\cdot \boldsymbol{\sigma} =k_x \sigma_x+k_y\sigma_y$,
$t_{soc}$ is the spin-orbit coupling strength, and $\mathbf{m}=(m_x,m_y,m_z)$ represents the magnetic moment.
The eigenvalues are $E_{i,\pm} = \pm [ k^2 + m_x^2 + m_z^2 + t_{soc}^2 + (-1)^i \times 2 \sqrt{ k^2 (m_x^2 + m_z^2) + m_z^2 t_{soc}^2}] ^{1/2}$. If one expands the eigenvalues up to $k^2$ order near $k=0$, the eigenvalues can be expressed as $E_{i,\pm} \approx \pm \sqrt{m_x^2+(m_z+(-1)^i t_{soc})^2} \mp[ k^2 (m_x^2+m_z(m_z+(-1)^i t_{soc}))]/[2 m_z t_{soc} \sqrt{m_x^2+(m_z+(-1)^i t_{soc})^2}]$. There is a band curvature reversion for $E_{1,\pm}$ by tuning $m_x,m_z$, and this reversion is the necessary condition for multi-Fermi surfaces of a single band.
\begin{figure}[tp!]
	\includegraphics[width=\columnwidth]{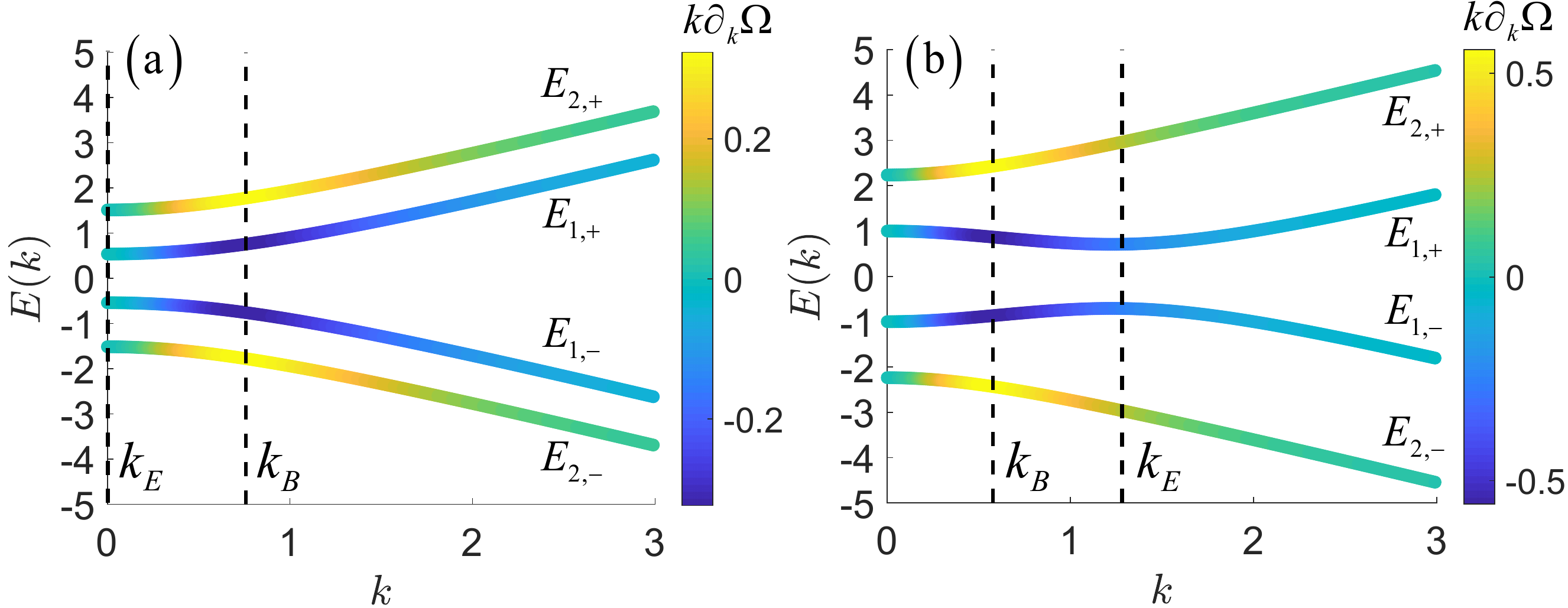}
	\caption{\textbf{Band dispersion and Fermi surface weight.} Band dispersion is indicated by colored lines and the Fermi surface weight $k\partial_{k} \Omega$ is labeled by the color bar ($t_{soc}=1$ for both figures). Dashed lines show the position $k_B$ where $\partial_{k}(k \partial_{k}\Omega)=0$ and $k_E$ where the minimal value of $E_{1,+}$. (a) For $m_x^2+m_z(m_z-t_{soc})<0$ case ($m_x=0.2,m_z=0.5$), $k=0$ is always the minimal point for $E_{1,+}$. (b) For $m_x^2+m_z^2>\sqrt{\frac{5}{3}} m_z t_{soc}$ case ($m_x=1,m_z=1$), $k_E>k_B$ and $k=k_E$ is always the minimal point for $E_{1,+}$ band.
	}
	\label{fig:fig1}
\end{figure}
Here we use $\pm$ in $E_{i,\pm}$ to label the band above/below 0 energy and $i$ to label the ith nearest band from 0 energy. The dispersion is plotted in Fig.~\ref{fig:fig1}. Berry curvatures for each band can be readily derived as
\begin{eqnarray}
\Omega_{i,\pm}=(-1)^{i+1} \frac{(m_x^2+m_z^2)m_z t_{soc}}{2 ((m_x^2+m_z^2)k^2+(m_z t_{soc})^2)^{3/2}}.
\label{eq:eq5}
\end{eqnarray}
By observing the results above, we can see the bands dispersion and Berry curvature distribution are isotropic, which are only determined by $k^2$ in momentum space, so that $m_x$ can point any in-plane direction without changing the conclusion. Besides, there is no band touching and reopening process at $k=0$ by tuning $m_x,m_z>0$ so that Berry curvature distribution is stable near $k=0$ point. Finally, $m_z<0$ case can be easily achieved by reversing Berry curvature of the $\left|  m_z\right| $ case. With these observations, we discuss the 3rd order AHE sign structure in this model below.

\noindent{\textcolor{blue}{\it The sign structure of the 3rd order AHE.}---} It is obvious that the sign of $t_{soc}$ or $m_z$ will change the sign of Berry curvature so that all the Hall signal will reverse (i.e. all linear and nonlinear Hall signal). This is the trivial sign switching by gap-close-reopen procedure. We will focus on the non-trivial sign switching mechanism which comes from multi-Fermi surface weighted sum in Eq.~\eqref{eq:eq3}, by setting $m_x,m_z,t_{soc}\geqslant0$ for convenience and finding the connection of the anisotropy magnetic moment with the sign structure of the 3rd order AHE. Because of the particle-hole symmetry of this model, we only consider the positive chemical potential.

Before analyzing the sign of the AHE, it is important to highlight one key feature of the band structure. Here, we focus on electron-doping with filling close to charge neutrality, i.e., only the band $E_{1,+}$ is partially filled in Fig.~\ref{fig:fig1}.
For $m_x^2+m_z(m_z-t_{soc})<0$, $E_{1,+}$ is a monotonic function of $k$ and thus if we set the chemical potential $\sqrt{m_x^2+(m_z-t_{soc})^2}<\mu<\sqrt{m_x^2+(m_z+t_{soc})^2}$ (i.e., we put the Fermi surface between minimal points of $E_{1,+}$ and $E_{2,+}$), the model system has only a single Fermi surface. According to  Eq.~\eqref{eq:eq3}, there is no sign change of the 3rd order anomalous Hall from single Fermi surface of single band $E_{1,+}$, if we vary $\mu$ or other control parameters. 

\begin{figure*}[htp]
	\includegraphics[width=0.8\textwidth]{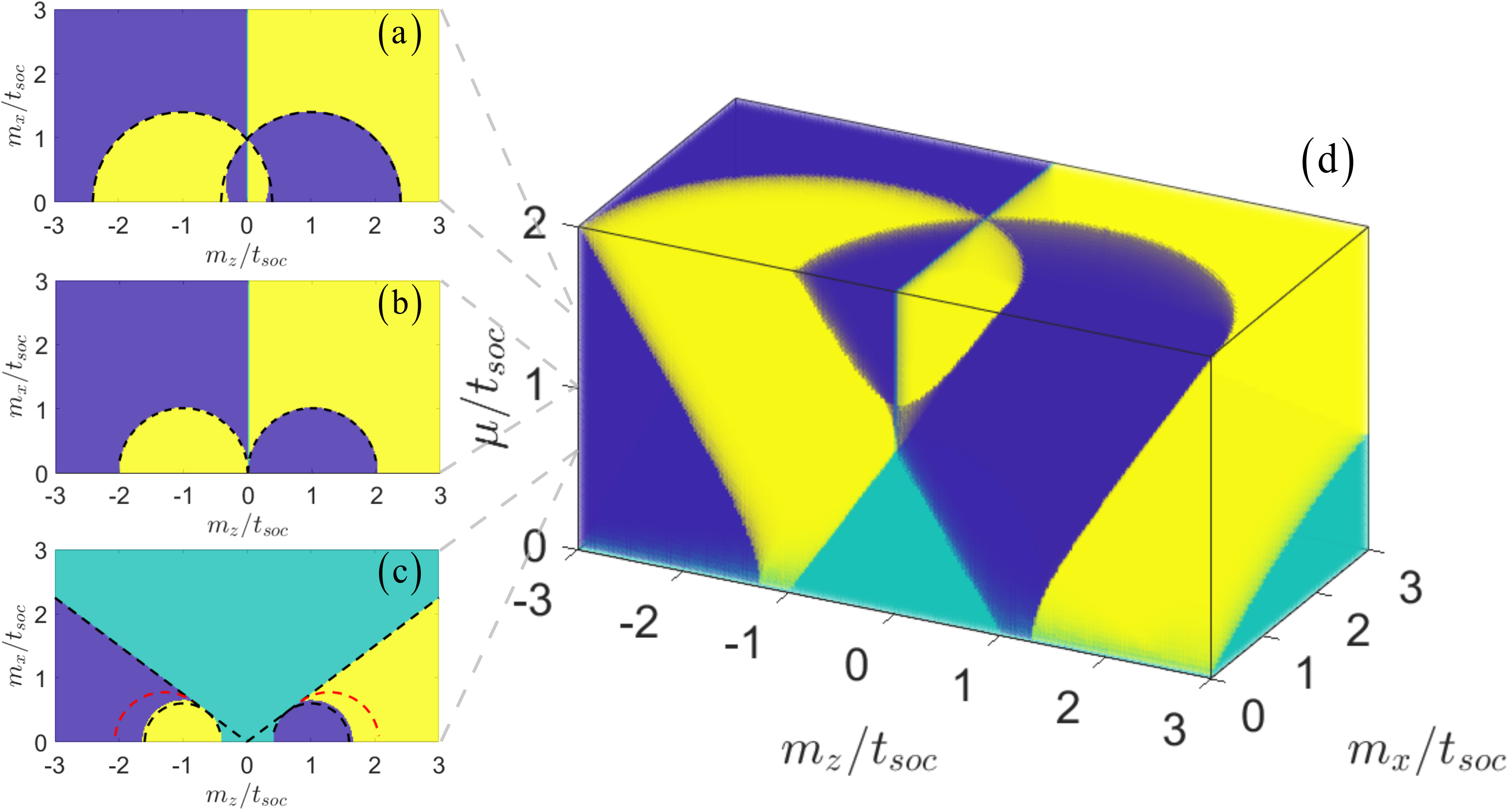}
	\caption{\textbf{Sign diagram of the 3rd order AHE.} The spin-orbital coupling strength is set as the unit of axis. The yellow region shows the positive sign and the blue region shows the negative. The cyan region means there is no signal. (a) For $\mu>t_{soc}$ interval, the chemical potential is set as $\mu=1.4 t_{soc}$. The left dashed semi-circle shows $\mu=\sqrt{m_x^2+(m_z+t_{soc})^2}$. Under this line, the $E_{2,+}$ band will contribute and finally change the sign from negative to positive when $m_z>0$. (b) For $\mu=t_{soc}=1$, there is no cyan region and the transition boundary forms a semi-circle $m_x^2+m_z^2=\sqrt{\frac{5 + 2 \sqrt{13}}{3}}m_z t_{soc}$ as shown by the black dashed lines. (c) For $0<\mu<t_{soc}$ interval, the chemical potential is set as $\mu=0.6 t_{soc}$. The straight dashed line shows $\mu=\frac{\left| m_x\right|  t_{soc}}{\sqrt{m_x^2+m_z^2}}$ and is the tangency of the black and red dashed semi-circles which indicate $\mu=\sqrt{m_x^2+(m_z\pm t_{soc})^2}$ and $\mu=\sqrt{\frac{3}{5}((m_z \pm \sqrt{\frac{5}{3}} t_{soc})^2 + m_x^2)}$. (d) Numerical results of the entire sign diagram of the 3rd order AHE when tuning $m_x, m_z, \mu$. The transition boundary approaches to two semi-cones. }
	\label{fig:fig2}
\end{figure*}

In contrast, for $m_x^2+m_z(m_z-t_{soc})>0$, the minimal value for $E_{1,+}$ is no longer located at $k=0$ point but at
\begin{equation}
	k_E=\frac{\sqrt{(m_x^2+m_z^2 )^2-m_z^2 t_{soc}^2}}{ \sqrt{m_x^2+m_z^2}},
	\label{eq:eq8}
\end{equation}
as denoted by the dashed line in Fig.~\ref{fig:fig1} (b). The minimal value of $E_{1,+}$ now is $\frac{m_x t_{soc}}{\sqrt{m_x^2+m_z^2}}$. With band structure, electron doping to the band $E_{1,+}$ will produce two Fermi surfaces, one with Fermi wave vector $k_F<k_E$ and the other $k_F>k_E$. These two Fermi surfaces have opposite Fermi velocity and thus can contribute opposite sign to $\sigma_{yxxx}^{(3\omega)}$. As shown in Fig.~\ref{fig:fig2}(d), the competition between these two Fermi surfaces gives rise to a very rich phase diagram with complicated sign structures.

Mathematically, the sign change boundary for $\sigma_{yxxx}^{(3\omega)}$ is the solution of equation
\begin{equation}
\sum_{n,s} sgn(\partial_{k} E_n) k \partial_{k} \Omega_n |_{k=k_{\mu,n,s}} = 0
\label{eq:eq6}
\end{equation}
Although the analytical solution is cumbersome, three simple and exact statements can be proved (see SI for the proof).

\noindent\textbf{Statement 1}
When $\sqrt{m_x^2+(m_z-t_{soc})^2}<\mu<\sqrt{m_x^2+(m_z+t_{soc})^2}$, there is no sign change of the 3rd order AHE when tuning $m_x, m_z, \mu$.

\noindent\textbf{Statement 2}
When $m_x^2+m_z^2>\sqrt{\frac{5}{3}} m_z t_{soc}$ and $\frac{m_x t_{soc}}{\sqrt{m_x^2+m_z^2}}<\mu<\sqrt{\frac{3}{5}((m_z - \sqrt{\frac{5}{3}} t_{soc})^2 + m_x^2)}$, there is no sign change of the 3rd order AHE by tuning the intensity of $m_x, m_z, \mu$. And the sign in this region is opposite to statement 1 region.

\noindent\textbf{Statement 3}
When $\mu=t_{soc}$, the sign switching boundary of the 3rd order AHE in $m_x, m_z$ plane is a semi-circle $m_x^2+m_z^2=\sqrt{\frac{5 + 2 \sqrt{13}}{3}}m_z t_{soc}$.

These exact conclusions dictate the global structure of the phase diagram, and in addition, they provide a pretty accurate estimation about the phase boundary. Here we plot the full phase digram [Fig.~\ref{fig:fig2}(d)], obtained numerically, and compares it with the exact statement above. For better visualization, in Fig.~\ref{fig:fig2}(a-c), we show the 2D cuts of the 3D phase diagram at three different chemical potential. For $0<\mu<t_{soc}$, the phase diagram is shown in the Fig.~\ref{fig:fig2}(c), where the yellow region shows the positive 3rd order AHE and the blue region shows the negative. The cyan region means there is no signal because of the chemical potential is below the minimum of $E_{1,+}$. While for the $\mu=t_{soc}$ case, one can see there is no cyan region from Fig.~\ref{fig:fig2}(b). The yellow region forms a semi-circle. For $\mu>t_{soc}$, one can see the constraint $\mu<\sqrt{m_x^2+(m_z+t_{soc})^2}$ will work when $m_z>0$ as the dash line from the other semi-circle shown in Fig.~\ref{fig:fig2}(a). Below this dashed line, $\mu>\sqrt{m_x^2+(m_z+t_{soc})^2}$ means Fermi surface from the $E_{2,+}$ band will contribute. Since the minimal point for $E_{2,+}$ band is always at $k=0$, the contribution from $E_{2,+}$ band is always positive so that there is a sign change from negative to positive below this line.


\begin{figure}[tp!]
	\includegraphics[width=1.0\columnwidth]{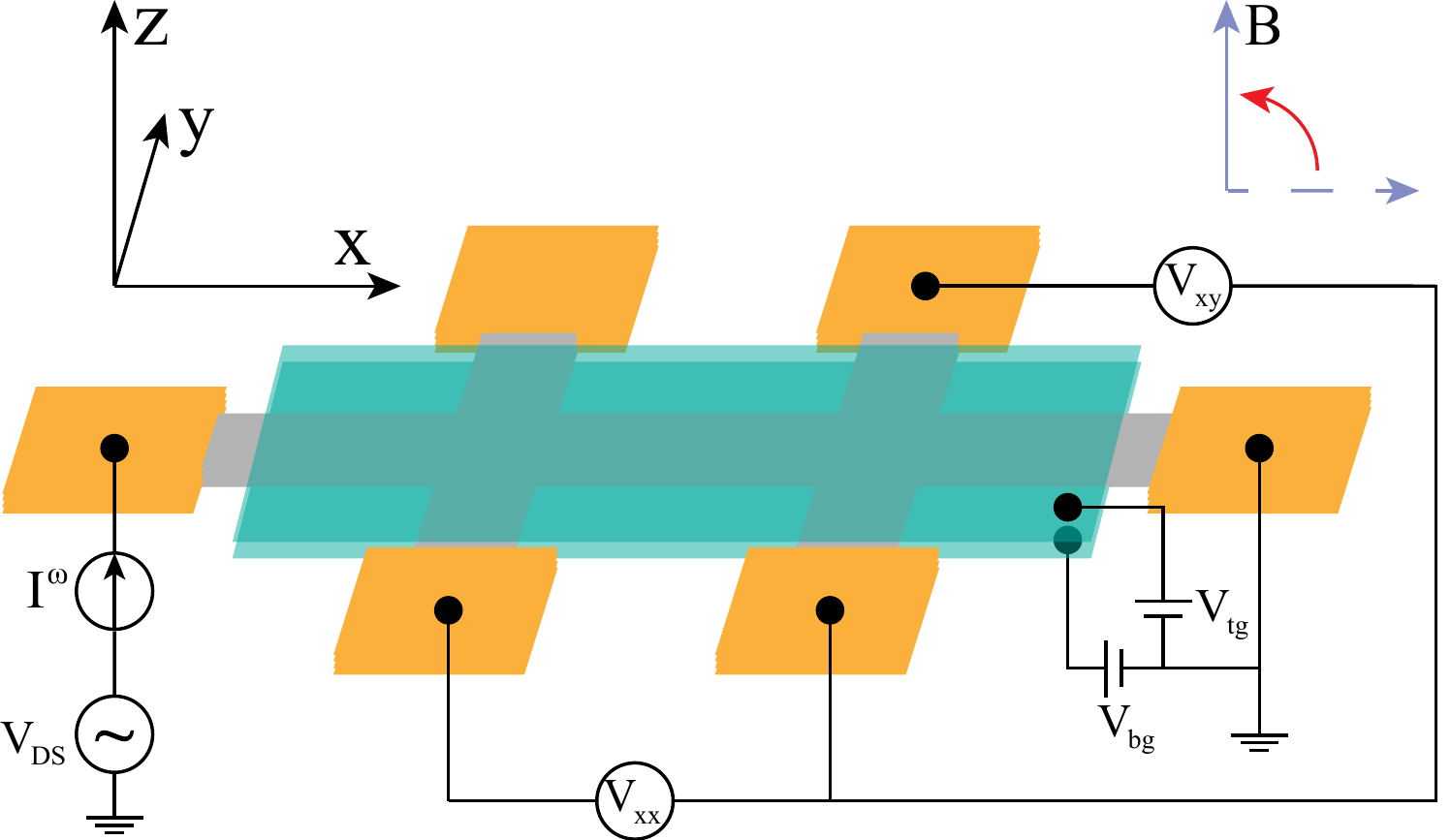}
	\caption{\textbf{Experimental detection for the sign switching in the 3rd order AHE.} Chemical potential is tuned by top gate and bottom gate voltage $V_{tg},V_{bg}$. AC current $I^{\omega}$ with low frequency $\omega$ injects along $x$ direction from source to drain. Then $V_{xx}$ and $V_{xy}$ for different frequency output can be measured. The magnetic anisotropy can be tuned by rotating the magnetic field from in-plane to out-of-plane directions.}
	\label{fig:fig3}
\end{figure}


\noindent{\textcolor{blue}{\it Experimental proposal and Discussions.}---} 
To detect the sign phase diagram of the 3rd order AHE, we propose the schematic experimental setting in Fig.~\ref{fig:fig3}. First, one requires a piece of clean 2D Dirac material with strong spin-orbital-coupling and large magnetic susceptibility. As the experiment of FeSn~\cite{berthold2023}, kagome metals satisfy this requirement. Silicene and germanene in group IVA also have a chance due to relative large spin-orbital-coupling, but a strong magnetic field is needed.
By forming a Hall bar geometry with double gate tuning the chemical potental and homogeneous magnetic field tuning magnetic moment, one can measure the anomalous contribution by subtracting the linear contribution $\rho_0^{(3\omega)} B$ from the 3rd order anomalous Hall resistivity $\rho_{xy}^{(3\omega)} (B), \rho_{A}^{(3\omega)} (B)=\rho_{xy}^{(3\omega)} (B)- \rho_0^{(3\omega)} B$ as shown in the experiment of FeSn~\cite{berthold2023}. By fixing the chemical potential and the strength of the magnetic field, rotating the magnetic field to change the magnetic anisotropy, the sign of $\rho_{A}^{(3\omega)} (B)$ shall change.

It is worthwhile to notice this sensitive response to magnetic anisotropy shall be the largest when chemical potential is tuned approaching to spin-orbital-coupling strength (i.e., $\mu\sim t_{soc}$ as shown in Fig.~\ref{fig:fig2}(b)), where for small $B$, any magnetization deviating from in-plane direction shall induce a sign change of $\rho_{A}^{(3\omega)} (B)$. This phenomena can be used to detect the spin-orbital-coupling strength $t_{soc}$ by observing the most sensitive $\mu$ point and also possible magnetic phase transition, as has been successfully demonstrated in Ref.~\cite{berthold2023}.

To conclude, we derive the 3rd order AHE formula in isotropic system as the weighted summation from each Fermi surface (Eq.~\eqref{eq:eq3}) and show this is the leading order where the sign of Fermi velocity $sgn(\partial_k E)$ contributes. To illustrate this point, we investigate the sign structure of the 3rd order AHE for a minimal model with spin-orbital coupling and magnetic anisotropy and plot the sign diagram with tuning magnetic moment $m_x,m_z$ and chemical potential $\mu$. Experiment for observing this phenomena is proposed, which can be used to detect spin-orbital-coupling strength of the material and magnetic anisotropy.

\noindent{\textcolor{blue}{\it Note added.}---}  We would also like to bring the readers attention to the experiment work from Berthold J\"ack's group appear on arXiv~\cite{berthold2023}, in which the 3rd anomalous Hall effect and its "sign problem" is observed at room temperature.

{\it Acknowledgements.---}We thank Berthold J\"ack and his group for sharing their experimental results. We thank Xuejian Gao, Chengping Zhang, Qi-Fang Li and Kam Tuen Law for inspiring discussions on the related topic. XZ and ZYM acknowledge support from the RGC of Hong Kong SAR of China (Grant Nos. 17301420, 17301721, AoE/P-701/20, 17309822), the ANR/RGC Joint Research Scheme sponsored by Research Grants Council of Hong Kong SAR of China and French National Reserach Agency(Porject No. A\_HKU703/22), XZ is funded in part by a QuantEmX grant from ICAM and the Gordon and Betty Moore Foundation through Grant GBMF9616 to Xu Zhang. We thank HPC2021 system under the Information Technology Services and the Blackbody HPC system at the Department of Physics, University of Hong Kong for providing computational resources that have contributed to the research results in this paper.
\bibliographystyle{apsrev4-2}
\bibliography{Dirac_Nonlinear}

\begin{appendix}
	\begin{widetext}
		\section{Explanation of statements}
		If the condition $m_x^2+m_z(m_z-t_{soc})<0$ holds true, one can see $E_{1,+},E_{2,+}$ bands only have their minimal values at $k=0$ point and monotonically increasing elsewhere, as shown in Fig.~\ref{fig:fig1} (a). When we consider one-band contribution, it means we require chemical potential $\sqrt{m_x^2+(m_z-t_{soc})^2}<\mu<\sqrt{m_x^2+(m_z+t_{soc})^2}$ (i.e., we put the Fermi surface between minimal points of $E_{1,+}$ and $E_{2,+}$). According to  Eq.~\eqref{eq:eq3}, there is no sign change of the 3rd order anomalous Hall from single Fermi surface of single band $E_{1,+}$. 
		
		If the condition becomes $m_x^2+m_z(m_z-t_{soc})>0$, one can see the minimal value for $E_{1,+}$ is not located at $k=0$ point but at another $k_E$ point
		\begin{equation}
		k_E=\frac{\sqrt{(m_x^2+m_z^2 )^2-m_z^2 t_{soc}^2}}{ \sqrt{m_x^2+m_z^2}},
		\label{eq:eq8}
		\end{equation}
		as denoted by the dashed line in Fig.~\ref{fig:fig1} (b). The minimal value of $E_{1,+}$ now is $\frac{m_x t_{soc}}{\sqrt{m_x^2+m_z^2}}$. Considering one-band contribution, we should require chemical potential $\frac{m_x t_{soc}}{\sqrt{m_x^2+m_z^2}}<\mu<\sqrt{m_x^2+(m_z+t_{soc})^2}$ now. In the subinterval $\sqrt{m_x^2+(m_z-t_{soc})^2}<\mu<\sqrt{m_x^2+(m_z+t_{soc})^2}$, one can see this is also a single band single Fermi surface contribution so that there is still no sign change. Combine these two intervals, we can form our first statement for our minimal model:
		
		\textbf{Statement 1}
		When $\sqrt{m_x^2+(m_z-t_{soc})^2}<\mu<\sqrt{m_x^2+(m_z+t_{soc})^2}$, there is no sign change of the 3rd order AHE when tuning $m_x, m_z, \mu$.
		
		While for the $\frac{m_x t_{soc}}{\sqrt{m_x^2+m_z^2}}<\mu<\sqrt{m_x^2+(m_z-t_{soc})^2}$ case, there are two Fermi surfaces with opposite $sgn(\partial_{k} E_{1,+})$ so that the competing may induce a sign change. We define $k_B$ as the extreme point for $k \partial_{k} \Omega_n$.
		\begin{equation}
		k_B=\sqrt{\frac{2}{3}} \frac{m_z t_{soc}}{ \sqrt{m_x^2+m_z^2}}.
		\label{eq:eq7}
		\end{equation}
		The $k_B$ momentum is denoted as the dashed line in Fig.~\ref{fig:fig1} and the energy $E_{1,+}$ at $k_B$ point is $E_B=\sqrt{(m_z - \sqrt{\frac{5}{3}} t_{soc})^2 + m_x^2 (1 - \frac{2 t_{soc}^2}{3(m_x^2 + m_z^2)})}$. It is obvious that when $k_E>k_B$, there must be at least two fillings where $\sigma_{yxxx}^{(3\omega)}$ is larger than zero (e.g., $\mu\sim\frac{m_x t_{soc}}{\sqrt{m_x^2+m_z^2}}$) and smaller than zero (e.g., $\mu\sim\sqrt{m_x^2+(m_z-t_{soc})^2}$). The $k_E>k_B$ condition is just $m_x^2+m_z^2>\sqrt{\frac{5}{3}} m_z t_{soc}$. It is also easy to derive when $k_E>k_B$ and $\mu<E_B$, the total weight of Fermi surfaces has the opposite sign with the statement 1 case. To identify this region easier, we release $\mu<E_B$ restriction a little according to $\frac{m_x t_{soc}}{\sqrt{m_x^2+m_z^2}}<\mu$ and require $\mu<\sqrt{\frac{3}{5}((m_z - \sqrt{\frac{5}{3}} t_{soc})^2 + m_x^2)}$. This leads us to another statement.
		
		\textbf{Statement 2}
		When $m_x^2+m_z^2>\sqrt{\frac{5}{3}} m_z t_{soc}$ and $\frac{m_x t_{soc}}{\sqrt{m_x^2+m_z^2}}<\mu<\sqrt{\frac{3}{5}((m_z - \sqrt{\frac{5}{3}} t_{soc})^2 + m_x^2)}$, there is no sign change of the 3rd order AHE by tuning the intensity of $m_x, m_z, \mu$. And the sign in this region is opposite to statement 1 region.
		
		It is also worth to note when $\mu=t_{soc}$, the shape of sign change boundary is just a perfect semi-circle. We conclude this as our statement 3, which can be derived directly by solving Eq.~\eqref{eq:eq6}.
		
		\textbf{Statement 3}
		When $\mu=t_{soc}$, the sign switching boundary of the 3rd order AHE in $m_x, m_z$ plane is a semi-circle $m_x^2+m_z^2=\sqrt{\frac{5 + 2 \sqrt{13}}{3}}m_z t_{soc}$.
	\end{widetext}
\end{appendix}

\end{document}